# Mechanism of Deep-focus Earthquakes Anomalous Statistics

V.D. Rusov[1*], V.N. Vaschenko[2], E.P. Linnik[1], S. C. Mavrodiev[3],
T.N. Zelentsova[1], L. Pintelina[1], V.P. Smolyar[1], L. Pekevski [4]

[1]*Odessa National Polytechnic University, Ukraine*,
[2]*National Antarctic Scientific Centre, Kiev, Ukraine*
[3]*Institute for Nuclear Research and Nuclear Energy, BAS, Sofia, Bulgaria,*
[4]*Seismological Observatory, Faculty of Natural Sciences and Mathematics, Skopje, Macedonia*

**Abstract**

Analyzing the NEIC-data we have shown that the spatial deep-focus earthquake distribution in the Earth interior over the 1993-2006 is characterized by the clearly defined periodical fine discrete structure with period $L$=50 km, which is solely generated by earthquakes with magnitude M$\in$[3.9; 5.3] and only on the convergent boundary of plates. To describe the formation of this structure we used the model of complex systems by A. Volynskii and S. Bazhenov. The key property of this model consists in the presence of a rigid coating on a soft substratum. It is shown that in subduction processes the role of a rigid coating plays the slab substance (lithosphere) and the upper mantle acts as a soft substratum. Within the framework of this model we have obtained the estimation of average values of stress in the upper mantle and Young's modulus for the oceanic slab (lithosphere) and upper mantle.

*Introduction.* Can the process of deformation of polymer film with a rigid coating be a model of geophysical processes? Answer to this, it would seem, the strange question is positive. Especially since it became clear that the structural and mechanical features of deformation of polymer film with a rigid coating reflect in fact the non-trivial physical properties of more general class of complex systems [1, 2]. In the literature the above complex systems are referred as "a rigid coating on a soft substratum" (RCSS) systems [1]. In such systems, the coating is an anisodiametric solid on the surface of stretched polymer-like film and undergoes uniaxial compressive deformation (Fig.1 (c)-(d)).

---

[*] Corresponding author: V.D. Rusov, e-mail siiis@te.net.ua

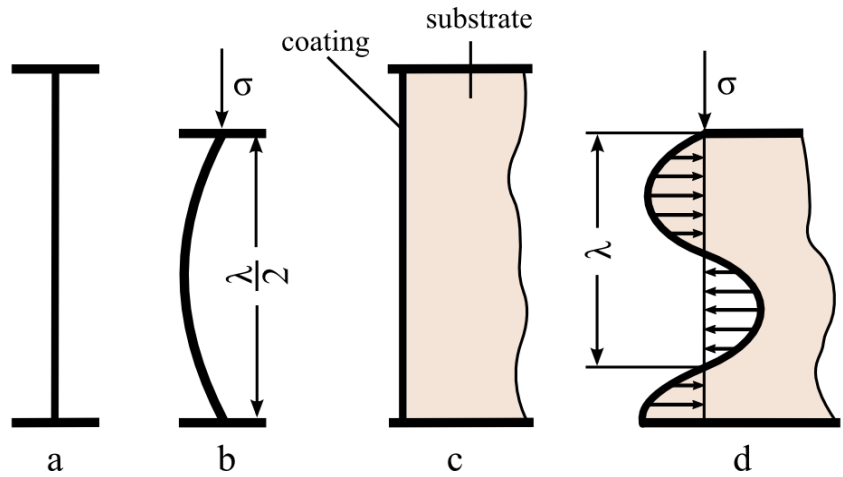

**FIG. 1.** Buckling instability for an anisodiametric body in a free state (a)-(b) and
on a soft substratum (c)-(d) [1].

Note that the task of elastic bar buckling (an anisodiametric solid) under uniaxial compression was first considered by Euler more than 200 years ago [3]. He has shown that a body buckles under the critical loading and acquires a half-wave form (Fig. 1 (a)-(b)). So-called Eulerian classical buckling was investigated later not only for bars but also for elastic plates, whose deformation, as assumed, was adequately described by the Karman equation [4], for example, in the form offered by Bauer and Reiss [5]. Eventually, just due to similar researches such fundamental concepts as universal deformation and structural stability have arose, which became later the fundamentals of dynamical system theory and catastrophe theory [6, 7].

On the other hand, if an anisodiametric solid (for example, a thin rigid coating) is firmly jointed with a resilient substrate, the picture of its buckling strikingly changes (Fig. 1 (c)-(d)). When the critical loading is attained, a body will not be able to acquire the half-wave form because at its deflection from the linear direction the restoring force from the substrate affects the body. This fore is proportional to the deviation value. As a result, such interaction of counteracting forces the coating is inevitably acquires a sinusoidal shape with the wavelength $\lambda$ (Fig. 1 (d)).

The problem of buckling and folding instability of sandwich system under compression was first considered by Smoluchowskii [8, 9] at the beginning of the twentieth century. Later this problem has been researched by Biot [10-12] and Allen [13]. In the case of an elastic layer lying on top of a semi-infinite elastic medium they showed, that a rigid coating bends under the action of compressive stresses with the formation of a regular surface relief. At the same time [10-13], the solution of equilibrium equation for sandwich systems under their loading, firstly, describes the sinusoidal wave pattern of coating and, secondly, makes it possible to finding the critical values of wavelength

$$\lambda = 2\pi h \left( \frac{(1-\nu_{sub}^2)E_{coat}}{3(1-\nu_{coat}^2)E_{sub}} \right)^{1/3} \quad (1)$$

and buckling stress in the coating

$$\sigma_{coat}^* = 0.52 \left( \frac{E_{coat}}{1-\nu_{coat}^2} \right)^{1/3} \left( \frac{E_{sub}}{1-\nu_{sub}^2} \right)^{2/3}, \quad (2)$$

where $h$ is the coating thickness, $E_{coat}$, $E_{sub}$ and $\nu_{coat}$, $\nu_{sub}$ are the Young moduli and the Poisson coefficients of the coating and substrate, respectively.

Now we consider the case when a substrate is soft (i.e., inelastic). This situation is typical just for the RCSS systems [1]. Eqs. (1)-(2) are formally valid in this case too, but there are some features related to the possible mechanism of regular fragmentation (cracking) of coating. For the first time it was shown in the modeling experiments by Ramberg and Stephanson [14], which could illustrate periodic of regular microrelief in the coating but failed to describe cracking in the coating, which also takes place in the RCSS systems under their loading. At the same time, cracking in the coating occurs in the following way. At fist cracks form in the antinodes of periodic regular micro relief and after that the mechanism of regular fragmentation (cracking) of coating is turned on when new-formed coating fragments break down into two fragments with equal dimensions. This mechanism of fracture via the breakdown of each fragment into two equal parts was first analytically analyzed in [15, 16]. As was shown, the average length $L$ of the fractured fragments along the drawing axis is equal to

$$L = 4h \frac{\sigma_{coat}^*}{\sigma_0}, \quad (3)$$

where $h$ is the coating thickness; $\sigma_{coat}^*$ is the ultimate strength of the coating; $\sigma_0$ is the stress in the support (the relief forming stress).

The offered approach is universal and can be applied not only for the estimation of ultrathin layers strength. All of structural features taking a place under deformation of films with coating are also inherent in other analogical systems. Results of researches [1] shown, that the formation of regular structures has a general nature and does not depend on a kind of substrate and coating. Necessary conditions determining the possibility of such structure formation due to deformation are the negligibly small coating thickness as compared with substrate thickness and the considerable difference between the modulus of elasticity (inflexibility) of the coating and substrate. In other words, the system must consist of inflexible thin solid resting on a soft pliable but thick substrate.

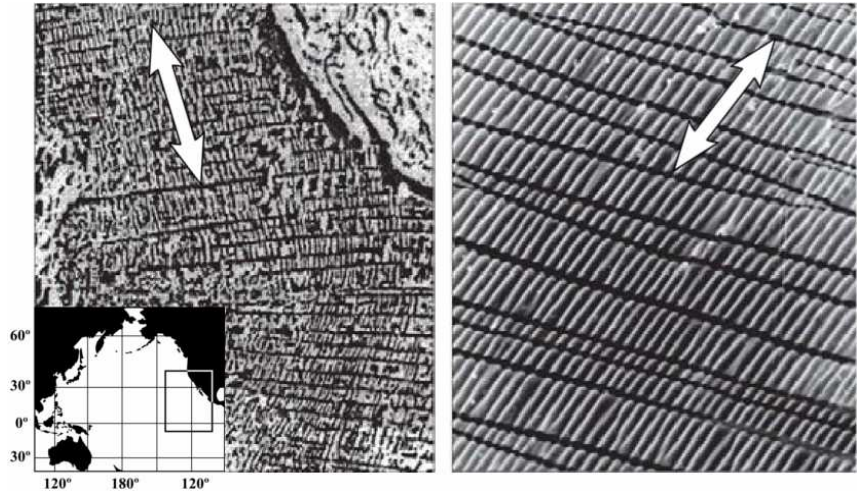

**FIG. 2.** The two similar reliefs of (a) the ocean floor in the region of the Eastern Pacific Rising (near the divergent boundaries of plates) and (b) gold coating which forms on the stretched rubber film [1]. White arrows show the tension direction. Insert - location of the ocean floor relief in the map of the Pacific Ocean.

Thus, since the above relationships appear to be universal [1], one may expect that they ay be applied for the description of the RCSS system of different physical nature. As has been noted by the authors of Ref. [1], such systems are very common in Nature, including fruit and bodies of animals. In their opinion, our planet Earth presents a gigantic RCSS system. For example, the surface micro relief in RCSS systems (Fig. 2(b)) is similar to the ocean floor relief in the vicinity of mid-oceanic ridges (Fig. 2(a)). Thus, the complex system composed of a young oceanic crust and the upper Earth's mantle may be considered as typically RCSS system, where a relatively thin (5-70 km) rigid external envelope of the Earth (lithosphere) rests on a relatively soft and thick (2900 km) envelope (upper mantle).

The importance of such RCSS system studies increases, if to take into account that a thermal or thermochemical convection in the upper mantle, which continuously generates mechanical stresses in the Earth's crust, is the reason of different fundamental geodynamic processes − continental drift, earthquakes, processes of relief formation. According to Ref. [1], analyze of such a system in term of the approach advanced for the description of the structural mechanical behavior of RCSS system, in particular, of the regular fragmentation of coating (Eq.(3)), allows not only to explain the processes of relief formation in the form of grid of transform fractures near convergent boundaries (or, in other words, the areas of spreading) but first to estimate the ultimate strength of the ocean floor [1].

In our opinion, the geosystem consisting of ocean slab (coating) which due to subduction sinks into the upper mantle (Fig. 3) and of the upper mantle (soft substrate) is one of the nontrivial RCSS geosystems. Therefore, the primary purpose of this paper is the finding of regular fragmentation mode in the ocean slab falling into the upper mantle at full subduction length up to the lower mantle roof (~700 km). The complementary but no less important task is the estimation of average value of stress in the upper mantle by the theory of regular fragmentation of the RCSS systems.

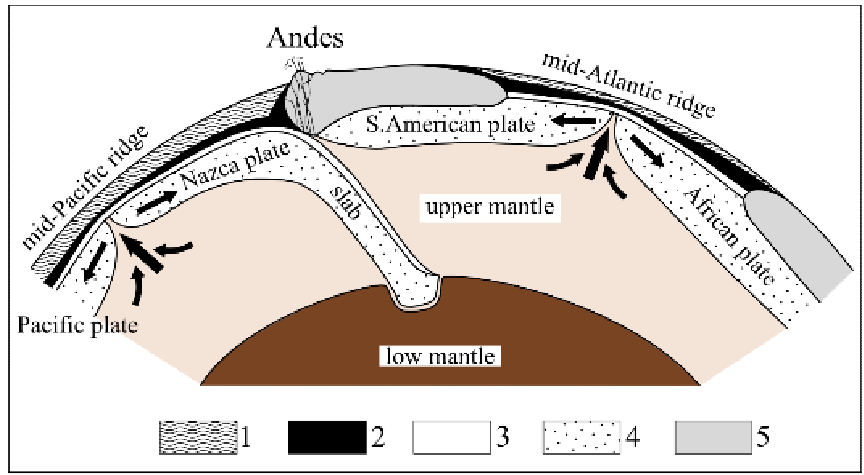

**FIG. 3.** The slab in the region of the Andes ridge in terms of the tectonic of the lithosphere plates:
1 - water; 2 - sediment; 3 - basaltic oceanic crust; 4 - lithosphere; 5 - continental crust.

**Regular fragmentation mode in the ocean slab and the fine periodical structure of deep-focus earthquakes**. It is obvious, that the ocean slab, which sinks into the upper mantle, is surrounded from every quarter by the mantle. Such a situation (Fig.4) generalizes the case of buckling instability for an aniso-diametric body on a soft substratum (Fig.1 (c)-(d)).

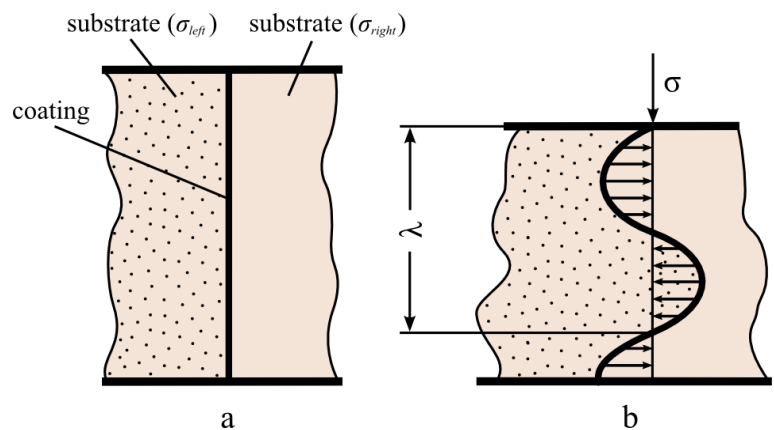

**FIG. 4.** The general case of buckling instability for an anisodiametric body surrounded from every quarter by soft substrata with different Young's moduli and Poisson's coefficients.

According to analytic analysis of the mechanism of regular fragmentation in the classical RCSS systems [15, 16] it is easy to show that in the general case (Fig. 4) the average length $L$ of the fractured fragments along the drawing axis is equal to

$$L = 4h \frac{\sigma^*_{coat}}{\sigma_{left} + \sigma_{right}} \approx 2h \frac{\sigma^*_{coat}}{\sigma_0}, \qquad (4)$$

where, ignoring a possible difference between the values of stress in the upper mantle to the left ($\sigma_{left}$) and to the right ($\sigma_{right}$) of slab (see Fig. 4) due to small difference in temperature, we will consider that $\sigma_{left} \approx \sigma_{right} \approx \sigma_0$.

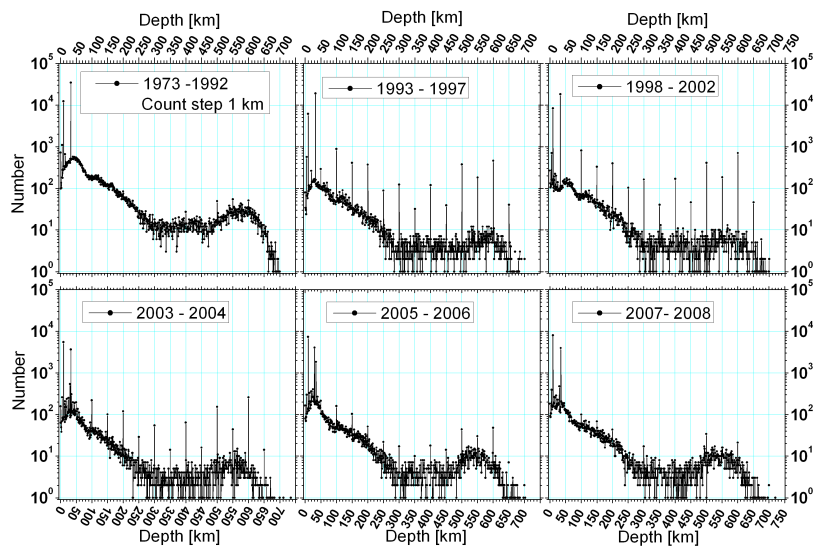

**FIG. 5.** Time evolution of depth distribution of deep-focus earthquakes with magnitude M≥3.9 over the 1973-2009 plotted on basis of the NEIC-data [17].

Using the NEIC data [17] we plotted the spatial deep-focus earthquake distribution in the Earth interior over the 1973-2009 (Fig. 5). From Fig. 5 follows that this distribution has clearly defined periodical fine discrete structure with period $L$=50 km. This structure is solely generated by earthquakes with magnitude M∈[3.9; 5.3] (Fig. 6) and only on the convergent boundary of plates (Fig. 7).

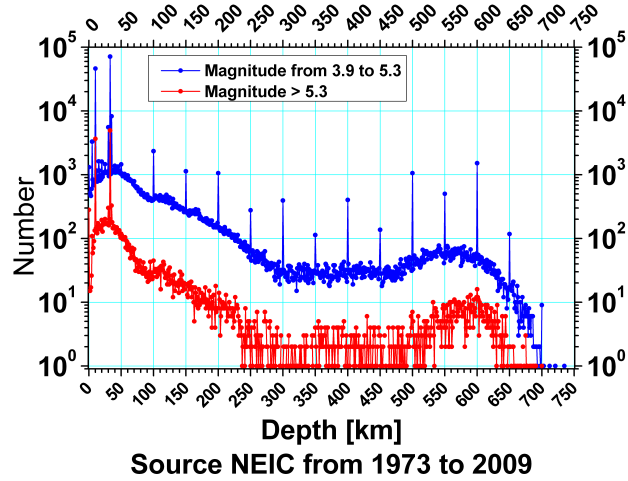

**FIG. 6.** The depth distributions of deep-focus earthquakes with magnitudes M∈[3.9, 5.3] and M≥5.3 over the 1973-2009 plotted on basis of the NEIC-data [17].

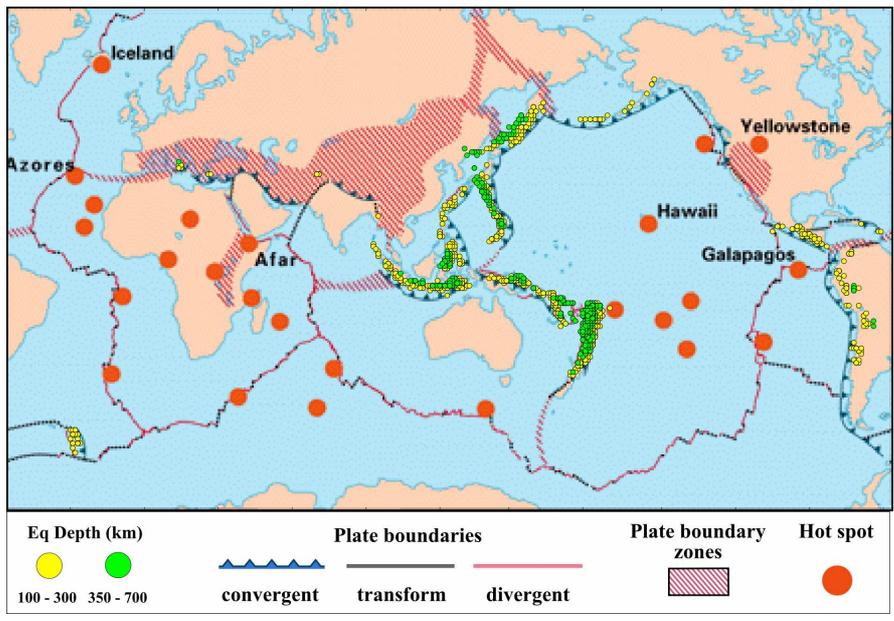

**FIG. 7.** The spatial distributions of deep-focus earthquakes with magnitude M∈[3.9, 5.3] owned to the fine structure of depth distributions of deep-focus earthquakes over the 1973-2006. The map is built on basis of Ref. [18].

Analysis of Figs. 5 and 6 shows that the process of slab cracking (lithosphere plate sunken into the mantle), which accompanies the chain of earthquakes, generates the two characteristic sizes of deep-focus earthquakes fine structure, i.e. 50 and 100 km. Supposing that earthquakes take a place, first of all, in the antinodes (maximum stresses) of sinusoidal wave pattern of slab, one of these sizes corresponds to the period $\lambda$=200 km of sinusoidal wave pattern of slab (see Eq. (1)) and is the corollary of folding instability of coating in the RCSS systems. Other size is equal the average length $L=50$ km of the fractured fragments and is the corollary of subsequent regular fragmentation of coating (see Eq. (4)). It should be noted that the slab cracking can become stronger due to the Coriolis force, which is especially effective in the antinodes (maximum stresses) of sinusoidal wave pattern of slab (see Fig. 4).

From now on we take into account that, according to seismography data for the different areas of subduction, the slab thickness decreases with the slab sinking into the upper mantle [19,20] and on the average becomes approximately equal to 1/2 from the initial (on entering the mantle) slab thickness. Then taking into account the known estimation of the ultimate strength of the lithosphere $\sigma^*_{coat} \sim 3$ MPa [1, 21] and supposing the effective slab thickness equal to $h \sim 15$ km, we obtain by Eq. (4) the estimation of stress in the upper mantle $\sigma_0 \sim 1.8$. MPa. It is interesting that the found value of stress in the upper mantle hierarchically (taking into account the gradients of temperature and density in the upper mantle) is in good agreement with the known estimation of stress $\sigma_0 \sim 0.6$ МПа, which sustains the convective motion (plastic flow) of mantel substance in an astenosphere [1].

Ignoring the small values of Poisson's coefficients for the slab and upper mantle and adopting that $\lambda=200$ km and $\sigma^*_{coat} \sim 3$ MPa, we obtain the following estimated values of Young's modulus for the slab and upper mantle $E_{slab} \sim 55$ MPa and $E_{mantle} \sim 1$ MPa by consistent solving of Eqs. (1) and (2). It will be recalled that Eqs. (1) и (2) in general case, i.e., for a layer of material embedded in an infinite medium (Fig. 4), are the same as for a layer of material lying on top of a semi-infinite medium (Fig.1 c-d) but with factor of 2 for Young's modulus $E_{sub}$. In other words, all results obtained for the a layer of material lying on top of a semi-infinite medium are applicable to the embedded layer if we replace $E_{sub}$ by 2 $E_{sub}$ [22].

*Summary.* Analyzing the NEIC-data we show that the spatial deep-focus earthquake distribution in the Earth interior over the 1993-2006 is characterized by the clearly defined periodical fine discrete structure with period $L=50$ km, which is solely generated by earthquakes with magnitude $M \in [3.9; 5.3]$ and only on the convergent boundary of plates. To describe the formation of this structure we used the model of RSCC systems, whose key property consists in the presence of a rigid coating on a soft substratum [1]. It is shown that in subduction processes the role of a rigid coating plays the slab substance (lithosphere plate) and the upper mantle acts as a soft substratum. Within the framework of this model we obtained the estimation of average values of stress in the upper mantle ($\sigma_0 \sim 1.8$ MPa) and Young's modulus for the lithosphere ($E_{slab} \sim 55$ MPa) and upper mantle ($E_{mantle} \sim 1$ MPa).

## References

[1]  A.L. Volynskii and S.I. Bazhenov, Eur. Phys. J. E **24**, 317 (2007).

[2]  N. Bowden et al., Nature **393**, 146 (1998).

[3]  L. Euler, *Methodus Inveniendi Lineus Curvus Maximi Minimive Proprietate Gaudetes (Appendix, De Curvus Elastics)* (Marcum Michaelum Bousquet, Lausanne and Geneva, 1744).

[4]  T. von Karman, Encyk. der Math. Wissen, **IV**, No.4, Teubner, Leipzig 348 (1910).

[5]  L. Bauer and E. Reiss, SIAM J. of Appl. Math. **13**, 603 (1965).

[6]  T. Poston and I. Stuart, *Catastrophe Theory and its Application* (Dover, New York, 1998).

[7]  V.I. Arnold, *Catastrophe Theory*, 3rd ed. (Springer–Verlag, Berlin, 1992).

[8]  M. Smoluchowskii, Abh. Acad. Wiss. Krakaw, Math. **K1**, 3 (1909).

[9]  M. Smoluchowskii, Abh. Acad. Wiss. Krakaw, Math. **K1**, 727 (1910).

[10]  M.A. Biot, J. Appl. Phys. **25,** 2133 (1954).

[11]  M.A. Biot, Q. Appl. Math. **17**, 722 (1959).

[12]  M.A. Biot, *Mechanics of Incremental Deformations* (Wiley, New York, 1965).

[13]  H.G. Allen *Analysis and Design of Structural Sandwich Panels* (Pergamon, NewYork, 1969).

[14]  H. Ramberg and O. Stephanson, Tectonophysics **1**, 101 (1964).

[15]  Y. Leterrier *et al.*, J. Polym. Sci. Part B: Polym. Phys. **35**, 1449 (1997).

[16]  A.L., Volynskii *et al.*, J. Appl. Polym. Sci. **72**, 1267 (1999).

[17]  National Earthquake Information Center – NEIC, USGS, Golden, Co. USA, http://earthquake.usgs.gov/regional/neic.

[18]  Source: http://pubs.usgs.gov/gip/dynamic/world_map.html.

[19]  K.L. Pankow and T. Lay, J. Geophys. Res. **104**, 7255 (1999).

[20]  L.L. Lobkovsky, A.M. Nikishin and V.E. Khain, *Current Problems of Geotectonics and Geodynamics* (Scientific World, Moscow, 2004).

[21]  C.P. Conrad and C. Lithgow-Bertelloni, Geophys. Res. Lett. **33**, L05312 (2006).

[22]  M.A. Biot, Proc. Roy. Soc. Lond., Ser. A **242**, 444 (1957).